\documentclass[conference]{IEEEtran}
\usepackage{tikz}
\usepackage{algorithm}
\usepackage{algpseudocode}
\usepackage{amsmath}
\usepackage{graphicx}
\usepackage{comment}
\usepackage{array,multirow,graphicx}
 \usepackage{float}
 \usepackage{amssymb}
 \usepackage{booktabs}
\usepackage{enumitem}
\usepackage{multicol}
\usepackage{makecell}
\usepackage{filecontents}
\usepackage{caption}  
\usepackage{subcaption}
\usepackage{longtable}
\usepackage{balance}
\algnewcommand\algorithmicforeach{\textbf{for each:}}
\algnewcommand\ForEach{\item[ \algorithmicforeach]}

\begin{document}

\title{ARA-O-RAN: End-to-End Programmable O-RAN Living Lab for Agriculture and Rural Communities}

 \author{\IEEEauthorblockN{Tianyi Zhang, Joshua Ofori Boateng, Taimoor UI Islam, Arsalan Ahmad, Hongwei Zhang, Daji Qiao}
 
 \IEEEauthorblockA{Department of Electrical and Computer Engineering, Iowa State University, U.S.A.\\
 \{tianyiz, jboateng, tislam, aahmad, hongwei, daji\}@iastate.edu
 }}
\maketitle
\thispagestyle{empty}

\begin{abstract}
As wireless networks evolve towards open architectures like O-RAN, testing, and integration platforms are crucial to address challenges like interoperability. This paper describes ARA-O-RAN, a novel O-RAN testbed established through the NSF Platforms for Advanced Wireless Research (PAWR) ARA platform. ARA provides an at-scale rural wireless living lab focused on technologies for digital agriculture and rural communities. As an O-RAN Alliance certified Open Testing and Integration Centre (OTIC), ARA launched ARA-O-RAN - the first public O-RAN testbed tailored to rural and agriculture use cases, together with the end-to-end, whole-stack programmability. ARA-O-RAN uniquely combines support for outdoor testing across a university campus, surrounding farmlands, and rural communities with a 50-node indoor sandbox. The testbed facilitates vital R\&D to implement open architectures that can meet rural connectivity needs. The paper outlines ARA-O-RAN's hardware system design, software architecture, and enabled research experiments. It also discusses plans aligned with national spectrum policy and rural spectrum innovation. ARA-O-RAN exemplifies the value of purpose-built wireless testbeds in accelerating impactful wireless research.

\end{abstract}




\maketitle
\section{INTRODUCTION} \label{sec:Introduction}
As wireless networking standards progress from 5G to Beyond 5G (B5G) and 6G, there is an increasing emphasis on realizing the vision of Open Radio Access Networks (O-RAN). O-RAN aims to revolutionize Radio Access Networks by embracing principles of openness, intelligence, virtualization, and vendor-agnostic interoperability. The merits of O-RAN encompass fostering market competition, offering customer choice, reducing equipment costs, and elevating network performance.

The O-RAN Alliance, founded in 2018, has the specific mandate to implement these O-RAN principles atop 3GPP LTE and NR RANs. This involves adopting and extending the 3GPP NR 7.2 functional split for base stations, which disaggregates base station functionalities into a Central Unit (CU), Distributed Unit (DU), and Radio Unit (RU). Furthermore, O-RAN introduces a novel software-defined component to its architecture in the form of two RAN Intelligent Controllers (RICs). Within the O-RAN architecture, the two types of RICs manage and control the network at near-real-time (10~ms to 1~s) and non-real-time (more than 1~s) time scales. These controllers play a pivotal role in the O-RAN disaggregation strategy by enabling multivendor interoperability, intelligence, agility, and programmability in RANs.

However, while the concept of a more 'open' RAN offers numerous advantages, it also presents challenges, as highlighted in \cite{SolvingT26:online}. These include ensuring interoperability, establishing ownership accountability, devising effective troubleshooting and isolation mechanisms, and managing the coexistence of multi-vendor virtual and physical network functions.

To address these challenges and expedite the R\&D progress of O-RAN, the O-RAN Alliance has introduced Open Testing and
Integration Centre (OTIC). OTICs play a crucial role in supporting the testing and integration of O-RAN technologies and products. In 2023, the Platforms for Advanced Wireless Research (PAWR) program's Advanced Wireless Research Platform for Digital Agriculture \& Rural Environments (ARA) received qualification from the O-RAN Alliance to host an OTIC~\cite{otic:online}.

ARA~\cite{zhang2022ara} as part of the National Science Foundation (NSF) PAWR program, is an at-scale platform for advanced wireless research deployed across the Iowa State University (ISU) campus, City of Ames, Iowa, USA, surrounding research and producer farms, and rural communities in central Iowa. ARA serves as a wireless living lab for smart and connected rural communities, facilitating the research and development of rural-focused wireless technologies that provide affordable, high-capacity connectivity to rural communities and industries such as agriculture. 

As an OTIC, ARA has chosen to enhance its commitment to O-RAN development by establishing a sub-lab for O-RAN testing within the ARA Wireless Living Lab. This new sub-lab, named ARA-O-RAN, is designed to serve as a testbed for researchers conducting O-RAN-related experiments.

It's worth noting that ARA-O-RAN is not the first public O-RAN testbed. In 2022, Powder introduced NexRAN, an open-source Open RAN use case within the POWDER framework~\cite{Johnson22NexRAN}. In the same year, the WIoT Institute of Northeastern University presented its OpenRAN Gym~\cite{bonati2022openran}. COSMOS~\cite{raychaudhuri2020cosmos} even took an earlier initiative by commencing the development of its COSMOS ONAP/ORAN Open Wireless Lab in 2019, although it has yet to claim O-RAN support on its public testbed. However, ARA-O-RAN boasts several unique attributes among O-RAN testbeds:

\begin{itemize}
\item It is the sole publicly available O-RAN testbed that facilitates end-to-end, whole-stack programmability in outdoor testing from the UEs to RUs, DUs, CUs, and core network. 
\item Positioned in a rural area and intricately integrated with agriculture, ARA-O-RAN stands as the only O-RAN living lab of its kind.
\item ARA-O-RAN offers a dedicated 50-node sandbox for O-RAN indoor testing. 
\end{itemize}
The subsequent sections of this paper are structured as follows: Section II outlines ARA-O-RAN hardware system design, while Section III delves into the software architecture. In Section IV, we provide a concise summary of the noteworthy O-RAN experiments conducted on ARA-O-RAN. Future plans are elucidated in Section V, and finally, Section VI presents the conclusion.

\section{SYSTEM DESIGN AND DEPLOYMENT ARCHITECTURE} \label{sec:Hardware}
\subsection{Deployment}\label{subsec:dep}
\begin{figure}[htbp]
    \centering
    \includegraphics[]{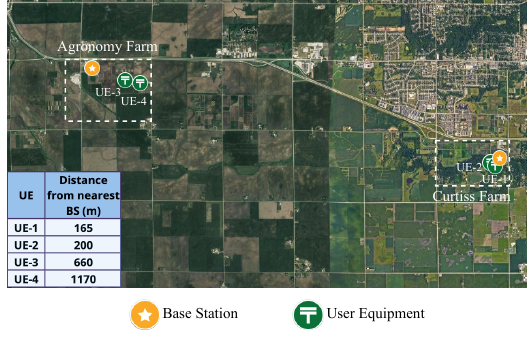}
    \caption{ARA Phase-1 Deployment Map.}
    \label{fig:map}
    \vspace*{-0.1in}
\end{figure}

The Phase-1 deployment of ARA is illustrated in Fig.~\ref{fig:map}, including four outdoor base stations (BSs) located in Ames, Iowa. Currently, O-RAN experiments are facilitated by two specific BSs, in Agronomy (Ag) Farm and Curtiss Farm, as denoted by the dashed rectangles in the deployment map. Notably, both Agronomy and Curtiss Farm function as active crop farms cultivating corn and beans. This unique setting offers users the opportunity to assess their algorithms and products in authentic wireless environments, with practical applications extending to areas such as agriculture automation.

\begin{figure}[htbp]
    \centering
    \includegraphics[scale=1.2]{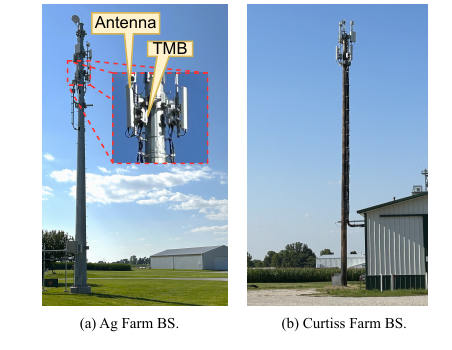}
    \caption{BSs at Ag Farm and Curtiss Farm.}
    \label{fig:field-ue}
    \vspace*{-0.2in}
\end{figure}

\begin{figure}[htbp]
    \centering
    \includegraphics[scale=0.95]{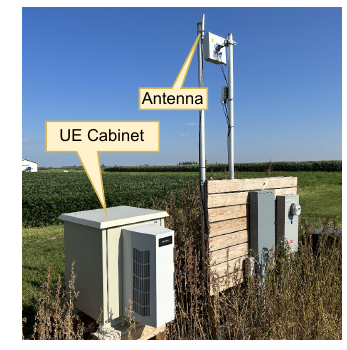}
    \caption{UE in the field.}
    \label{fig:UE}
    \vspace*{-0.25in}
\end{figure}

Within the coverage area of the BSs, there are 10 fixed-location UEs strategically placed to facilitate comprehensive wireless research. Additionally, there are also mobile UEs deployed on agricultural vehicles. Specifically, the BSs deployed at Ag farm and Curtiss farm are illustrated in Figure \ref{fig:field-ue}, and one field UE is shown in Figure \ref{fig:UE}. Each base station cell ensures a minimum of two UEs within its working range. Presently, for O-RAN experiments, we have 4 UE nodes (2 in Agronomy Farm and 2 in Curtiss Farm) in operational readiness.

The ARA team is actively engaged in building Phase-2, aiming to incorporate additional BSs and UEs to support diverse wireless research, including O-RAN. Notably, mobile UEs installed on buses, fire commander vehicles, and agriculture vehicles are currently in the deployment phase, with an expected availability by May 2024. These mobile UEs will play a crucial role in supporting research and experimentation in dynamic, mobile environments.

\begin{figure}[htbp]
    \centering
    \includegraphics[width=\columnwidth]{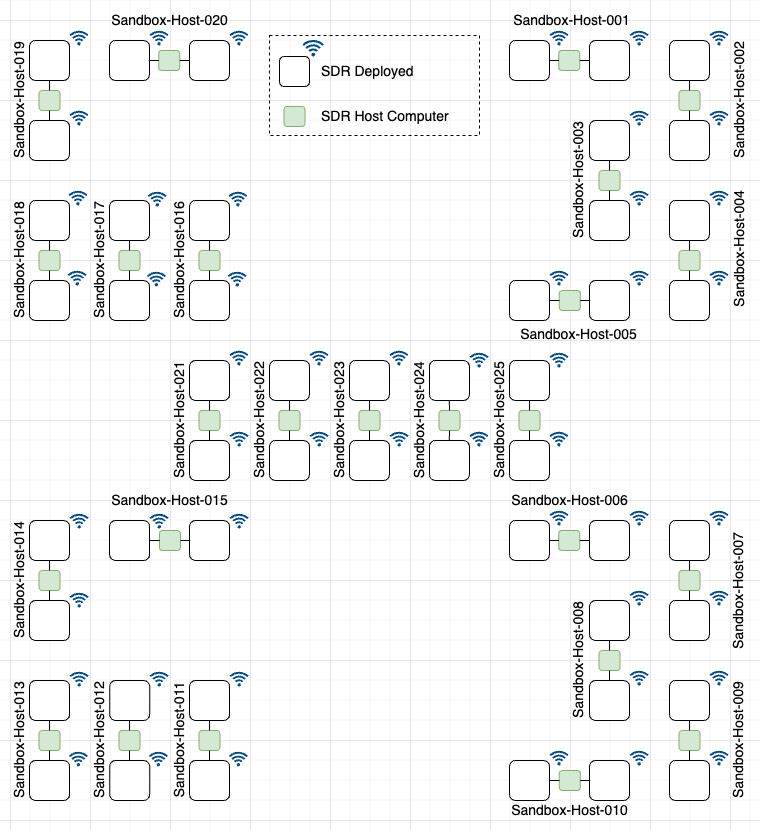}
    \caption{ARA Sandbox Topology.}
    \label{fig:Sandbox}
    \vspace*{-0.2in}
\end{figure}

In addition to outdoor deployment, ARA offers a controlled sandbox environment for users to conduct experiments in a laboratory setting with more nodes available. The sandbox features 50 USRP B210s arranged within a room and strategically mounted on a panel just below the room's ceiling. The mounting topology is illustrated in the Fig.~\ref{fig:Sandbox}. Each host computer is linked to two SDRs and the all the computers establish connectivity with the ARA cloud via switches.

\subsection{Base Station Design}

Each ARA BS contains three groups of devices, with each group covering a 120-degree sector. These groups are equipped with a comprehensive array of wireless equipment to support diverse testing purposes, including mmWave and MIMO. For O-RAN experiments, the primary reliance is on USRP N320 software-defined radios (SDRs) for transmitting and receiving signals. Combining with the open-source software, N320 SDRs grant users the flexibility to modify existing 5G/O-RAN code, facilitating the development and testing of ideas and solutions within a generous bandwidth of up to 200 MHz.

USRP SDRs, designed for indoor use, exhibit insufficient signal strength in outdoor settings. To address this limitation, ARA BSs incorporate tower-mounted boosters (TMBs) positioned between the antennas and SDRs to amplify both transmitting and receiving signals. The TMB is composed of a power amplifier (PA) and a low-noise amplifier (LNA), amplifying signals from TX and to RX ports separately. Further details can be explored in \cite{unpublishedkeyB}. With the assistance of TMBs, the coverage distance of ARA BSs extends beyond 1 mile, enabling outdoor O-RAN experiments.

All ARA base stations (BSs) establish connectivity to ARA servers via AraHaul, a distinctive x-haul system designed to accommodate both conventional backhaul services and research studies related to x-haul\cite{zuarahaul}. This interconnection is facilitated through the fiber or wireless backhaul transmission. The ARA servers play a dual role by furnishing open-source core networks essential for O-RAN experiments and facilitating remote access to the BSs. This remote accessibility empowers users to conduct O-RAN experiments, gather data, and allows ARA to efficiently manage and update the BS system in real-time.

\subsection{User Equipment Design}
Similar to the BSs, each UE box in the ARA infrastructure is outfitted with a cluster of devices to support diverse experiments. For O-RAN experiments, the setup employs USRP B210 SDRs in conjunction with User Equipment Boosters (UEBs), akin to a smaller version of the TMBs. Like their larger counterparts, the primary function of UEBs is to amplify signals, enabling the utilization of SDRs in outdoor, long-distance wireless communications scenarios.

To maintain wireless connections, all UE boxes utilize commercial devices as management channels. These channels enable users to remotely control the UE nodes, facilitating experimentation and providing a mechanism for users to conduct experiments effectively.

\subsection{Additional Resources}\label{subsec:AR}
Functioning as an outdoor wireless living lab, ARA extends additional resources to assist users in gathering diverse datasets to enhance their experiments. For instance, ARA features weather stations strategically positioned near the antennas of the BSs, offering high-precision, firsthand weather information complete with timestamps. Furthermore, both the BS and UE nodes are equipped with extra SDRs for spectrum monitoring. These supplementary resources empower users to capture meaningful data, thereby enriching their O-RAN research endeavors. More details of available resources and APIs can be found in ARA User Manual~\cite{ARA-user-manual}.

\section{SOFTWARE ARCHITECTURE} \label{sec:Software}
\subsection{Software Framework}
\begin{figure*}[ht!]
  \centering
  \includegraphics[scale=0.99]{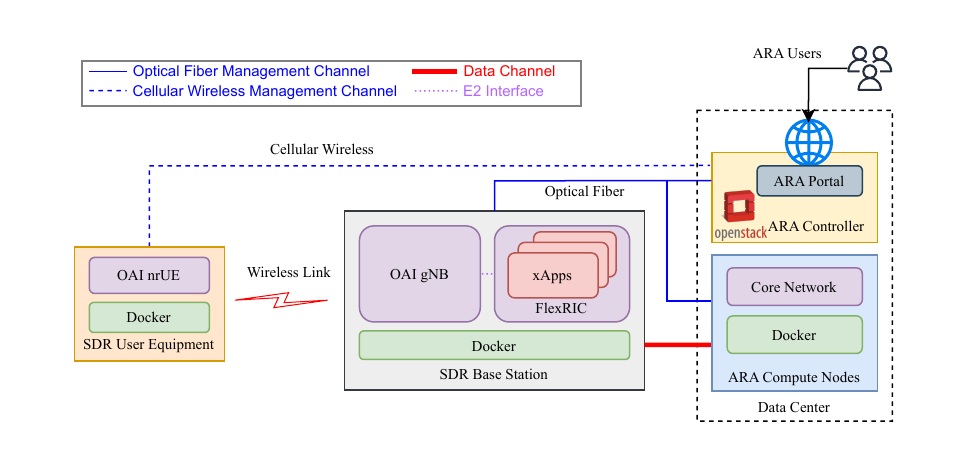}
  \caption{ARA-O-RAN Software Architecture.}
  \label{fig:arasoft_arch}
  \vspace*{-0.2in}
\end{figure*}

ARA software framework is derived from the OpenStack cloud operating system with container-mode resource provisioning, the philosophy adopted by the CHI@Edge spin of Chameleon Cloud~\cite{keahey2020lessons}. In addition to conventional resources such as compute and storage, ARA introduces wireless resources into the cloud platform, empowering researchers to conduct advanced wireless research. The foundational software architecture of ARA-O-RAN is outlined in Fig. \ref{fig:arasoft_arch}.

ARA's data center serves two primary functions. Firstly, it provides remote access for users. Upon registration, users can log into the ARA Portal and create reservations for both computer and wireless resources. Given that all nodes are container-based, users must claim the Docker images they wish to use during experiments. While users have the option to upload and utilize their own Docker images, ARA also provides pre-built Docker images for various experiments, including those related to O-RAN.

For O-RAN experiments, the BS Docker image encompasses the open-source 5G software, OpenAirInterface (OAI)~\cite{8727207}, along with the open-source near-Real-Time RAN Intelligent Controller (near-RT-RIC), flexRIC. These components come pre-configured and ready for use, only requiring users to open three terminals to independently run the gNB, near-RT-RIC, and xApps. Comprehensive details on utilizing pre-built Docker images can be found in the ARA user manual~\cite{ARA-user-manual}. Users are also encouraged to download images from the publicly available DockerHub, enabling them to develop their images based on the pre-built ones for testing within the ARA testbed. In the case of Docker images for UEs, ARA employs a pre-built image of OAI nrUE.

Upon the commencement of the reservation period, the ARA controller generates containers based on user requests through the management channels mentioned in Section~\ref{sec:Hardware}, utilizing images stored in the ARA data center. Once containers are ready, users gain access to their reserved nodes and can initiate experiments concurrently with the allocated radio resources.

The workflow for utilizing ARA-O-RAN simplifies as follows:
\begin{enumerate}
\item A user logs into the ARA Portal and creates a reservation for at least 1 BS node and 1 UE node.
\item The user selects either a pre-built Docker image or uploads their own Docker image.
\item The user launches the containers.
\item The user initiates the gNB, Near-RT-RIC, and xApps in the BS node, along with the nrUE in the UE node, to commence experiments.
\end{enumerate}

Currently, ARA-O-RAN facilitates experiments involving near-RT-RIC, in conjunction with other essential 5G components such as gNB, UE, and core networks.  A detailed tutorial on conducting O-RAN experiments within the ARA platform is provided in \cite{ARA-user-manual}. Fully-packaged experiments with open-source Service Management and Orchestration (SMO) and non-RT-RIC are expected to be made available in the near future.

\subsection{OpenAirInterface}
OpenAirInterface~\cite{8727207} is an open-source software implementation of the LTE/5G system, encompassing the entire protocol stack specified by the 3GPP standard. In the ARA environment, a specialized version of the OAI gNB with E2 interface is employed to establish connectivity with the near-RT-RIC. The choice of OAI is driven by its status as a comprehensive, full-stack open-source software, providing users the flexibility to conduct end-to-end experiments with the ability to customize nearly every aspect. Within ARA, both the gNB and nrUE are deployed within Docker containers, facilitating experiment reproducibility and offering flexibility for experimenters to utilize their pre-built Docker container images. While other open-source 5G software, such as srsRAN, exists, it is currently not supported in ARA-O-RAN. Although alternative open-source 5G software options, such as srsRAN, do exist, they are currently not supported in ARA-O-RAN, but it's worth noting that srsRAN has successfully passed internal testing and is slated to be made available to the public in the near future.

\subsection{FlexRIC}
FlexRIC~\cite{schmidt2021flexric} is an open-source near-RT-RIC developed by EURECOM. Positioned as a software-defined component within the O-RAN architecture, it is tasked with controlling and optimizing RAN functions. The near-RT-RIC control RAN elements and their resources, executing optimization actions typically within the time frame of 10 milliseconds to one second. It also receives policy guidance from the non-RT RIC and provides policy feedback to the non-RT-RIC through specialized applications known as xApps.

Given that both FlexRIC and OAI are products of the same organization, FlexRIC seamlessly integrates with OAI compared to other open-source near-RT-RICs. To streamline the onboarding process for new O-RAN researchers and mitigate potential challenges arising from dependency package updates, ARA has created a Docker image featuring a stable version of FlexRIC, alongside OAI, stored in the ARA Data Center. This initiative aims to save researchers time by eliminating the need to prepare the environment and install FlexRIC.

xApps are applications designed to run on the near-RT-RIC to manage network functions in near-real time. These applications are independent of the Near-RT RIC and can be provided by any third party. In FlexRIC, users have the capability to develop their xApps in both C and Python. Tutorials on creating xApps in FlexRIC can be accessed at \cite{CreatexA34:online}.

\section{ENABLED O-RAN EXPERIMENTS} \label{sec:Experiment}
In this section, we provide an overview of the featured O-RAN experiments supported by ARA. Due to space constraints, we present a selection of unique experiments offered by ARA, along with those deemed beneficial for researchers.
\subsection{Outdoor Experiment under Real-World Scenarios}

In contrast to simulations and indoor labs, ARA-O-RAN offers a distinctive opportunity for users to conduct O-RAN experiments in real-world agriculture and rural community settings. Leveraging the BSs and UEs situated in Crop Farms, users can perform outdoor experiments, enabling the evaluation of algorithms and products in authentic wireless environments.

Moreover, ARA extends an opportunity for users to gather performance measurement results by testing their solutions within real-world applications. Serving not only as a wireless research testbed for technological exploration but also as a provider of wireless connection services for agriculture projects, ARA supports modern agriculture technologies such as digital and precision agriculture. These applications often demand real-time high-throughput video streaming, and ARA users can assess their solutions in such scenarios, facilitating a comprehensive analysis.

Furthermore, as highlighted in Section~\ref{subsec:AR}, during outdoor experiments, users can conduct quantitative analyses of the weather impact and spectrum occupation by leveraging additional resources available in the outdoor environment. This multifaceted approach allows researchers to gain insights into the performance of their solutions under diverse real-world conditions, contributing to a more thorough understanding of the system dynamics.

\subsection{Bring-your-own-device (BYOD) Experiment}
As an OTIC, ARA is committed to expediting O-RAN-related R\&D by supporting BYOD experiments. In preparation for on-site testing, during the design and construction of the BSs and UEs, additional spaces, power sockets, and ports were intentionally incorporated to accommodate the integration of extra devices into the ARA system. To date, ARA facilitates users in testing their near-RT-RIC by establishing remote connections with ARA gNBs. Furthermore, ARA is in the process of creating a new testing environment featuring separate containers for O-CU, O-DU, Core network, and near-RT-RIC, with open F1 and E2 interfaces in the ARA network. This initiative will empower users to test various O-RAN components, including O-RU, O-DU, and O-CU, by simply connecting them to the ARA network with the correct interface IP addresses.

\subsection{Large-scale Experiment}
While indoor experiments may not hold the same allure as their outdoor counterparts, within ARA, the sandbox offers unparalleled advantages in terms of scalability when compared to the outdoor testbed. As discussed in Section~\ref{subsec:dep}, the ARA sandbox boasts 50 SDRs, all available for O-RAN indoor experiments. In contrast to the limited number of outdoor nodes in ARA and other public testbeds, this substantial resource pool of 50 nodes enables the establishment of a large network for testing interactions between nodes. For instance, a typical near-RT-RIC often connects to multiple O-CUs, O-DUs, and O-eNBs. Introducing interfaces like X2, Xn, and NG between different O-CUs for exchanging information further complicates the network dynamics. These new interfaces and intricate connections pose challenges for near-RT-RIC design, and the most effective way to address these challenges is through comprehensive testing in a large-scale, real-world testbed rather than relying solely on simulations.

\subsection{End-to-end, Whole-stack Programmability Experiment}
ARA places a primary emphasis on the end-to-end, whole-stack programmability of its testbed right from the outset. While some algorithms can be easily implemented on one RAN component without necessitating changes in others, numerous algorithms require comprehensive adjustments. For instance, machine learning applications heavily rely on training data and ARA's programmable UEs enable the design of new protocols, allowing UEs to gather additional data for the training of xApps in near-RT-RICs.

Leveraging its end-to-end, fully programmable SDR infrastructure and the advent of open-source O-RAN platforms such as OpenAirInterface and srsRAN, ARA facilitates comprehensive end-to-end programmability for O-RAN experiments. This capability extends to every element within the system, encompassing UEs, RU, DU, CU, RICs, and the core network. ARA's testbed enables programming and testing in real-world environmental and application settings, empowering researchers to explore the full spectrum of possibilities.

\subsection{Near-RT-RIC and xApp Experiment}
Near-RT-RICs and xApps have garnered substantial attention from both industries and academia for their potential to infuse intelligence into RANs. However, not all researchers interested in O-RAN or RICs are well-versed in open-source 5G/O-RAN platforms like OAI. Consequently, ARA's container-based testing platform offers an accessible entry point, ensuring a seamless initiation into this domain.

\begin{figure}[htbp]
    \centering
    \includegraphics[width=0.9\columnwidth]{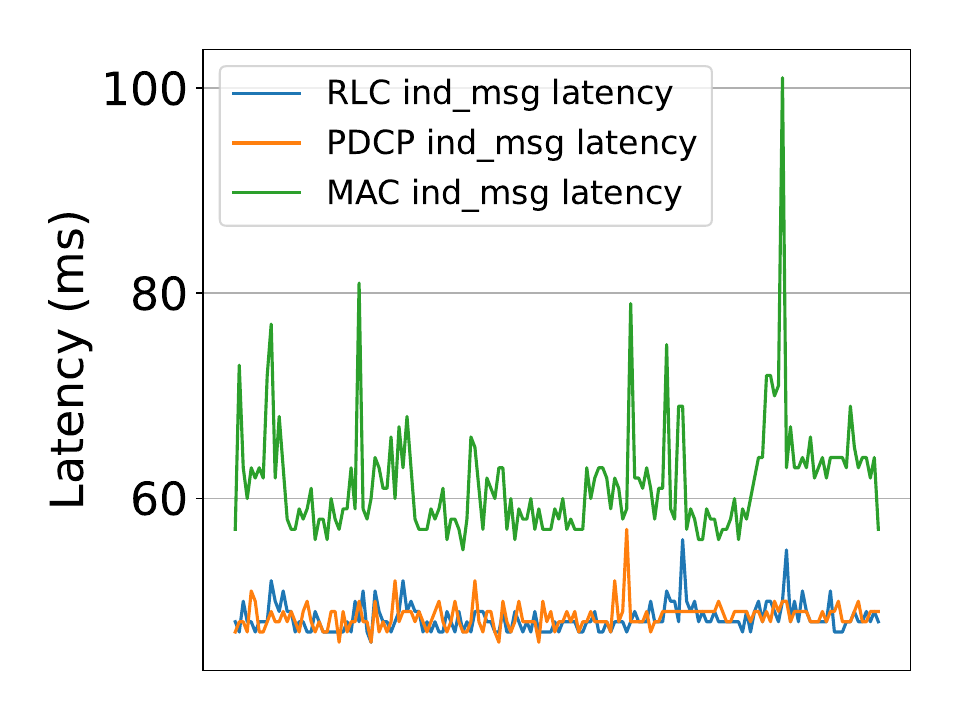}
    \caption{The indication message latency (ms) of RLC/DPCP/MAC over time.}
    \label{fig:line}
    \vspace*{-0.2in}
\end{figure}

For new users, ARA provides an uncomplicated start by offering pre-built, ready-to-use container images. This eliminates the need for spending days or even weeks on building the environment and systems from scratch. ARA's tutorials~\cite{ARA-user-manual} guide users through the process, enabling them to swiftly learn to run FlexRIC alongside OAI. Particularly beneficial for researchers focusing solely on xApps, ARA supports the direct development of xApps in Python/C, allowing users to efficiently test their applications within the ARA platform. This streamlined approach ensures a painless onboarding experience, encouraging broader participation and exploration within the O-RAN landscape. As a simple example, Fig. \ref{fig:line} shows the latency information collected by one xApp.

\section{FUTURE PLANs} \label{sec:Future}
In addition to the previously outlined initiatives such as expanding the deployment with more BSs and UEs, and incorporating support for srsRAN, ARA-O-RAN's principal objective revolves around enhancing the testing capabilities for various O-RAN components. In pursuit of this goal, ARA is concurrently advancing through two key avenues:
\begin{itemize}
    
    \item Open-Source Integration: ARA is actively working on integrating additional open-source O-RAN components into ARA-O-RAN, with a special focus on components available from the O-RAN Software Community. This collaborative effort aims to broaden the spectrum of O-RAN elements available for experimentation within the ARA platform.
    \item Commercial Collaboration: ARA is engaged in collaborative efforts with leading testing equipment vendors such as Keysight and Viavi. The primary objective is to formulate a robust testing solution that not only offers a comprehensive and professional O-RAN testing environment but also strives to preserve the programmability inherent in open-source solutions. In addition to the commercial testing equipment, ARA also hosts production-grade TVWS programmable mMIMO deployment \cite{aramimo-wintech2023} which follows CU, DU and RU architecture that will become O-RAN compliant in near future, enabling Open-RAN mMIMO studies at PHY/MAC layer, where real-time and near-real-time RIC are crucial for making scheduling decisions under MU-MIMO settings. 
    
\end{itemize}
This dual approach seeks to provide users with a well-rounded testing ecosystem that balances the benefits of open-source flexibility with the depth of commercial testing expertise.

\section{CONCLUSION} \label{sec:Conclusion}
In this paper, we introduce ARA-O-RAN, an end-to-end whole-stack programmable outdoor O-RAN living lab. As a integral part of the ARA wireless testbed, this newly available public lab welcomes new users and is eager to support their research by providing a real-world O-RAN experimentation platform.

Compared to other O-RAN testbeds, ARA-O-RAN holds a unique advantage of offering a real-world testing environment, allowing users to gather measurement results by testing their solutions within agricultural applications. ARA-O-RAN also supports various other experiments, including BYOD experiments, large-scale indoor experiments, and xApps development.

Our intention with this paper is to introduce ARA-O-RAN to potential researchers interested in O-RAN, with the hope of accelerating the development of O-RAN.

\section*{Acknowledgment}

\vspace{-0.2cm}

This work is supported in part by the NIFA award 2021-67021-33775, and NSF awards 2130889, 2112606, 2212573, 2229654, 2232461.

\vspace{-0.4cm}



\balance

\bibliographystyle{IEEEtran}
\bibliography{references}

\end{document}